\documentclass[12pt,a4paper]{article}
\usepackage[T1]{fontenc}
\usepackage[utf8]{inputenc}
\usepackage[american]{babel}
\usepackage{lmodern,microtype,amsmath,amssymb,amsthm,booktabs,array}
\usepackage[margin=30mm]{geometry}
\usepackage[authoryear,round]{natbib}
\usepackage{xurl,xcolor}
\usepackage[colorlinks=true,linkcolor=blue!45!black,citecolor=blue!45!black,urlcolor=blue!45!black]{hyperref}
\newtheorem{proposition}{Proposition}
\newtheorem{lemma}{Lemma}
\newcommand{\vb}{\overline v}
\newcommand{\E}{\mathcal E}
\newcommand{\loss}{\mathcal L}
\hypersetup{pdftitle={Credit Access and Mutual Risk Sharing},pdfauthor={Georgy Lukyanov}}
\title{Credit Access and Mutual Risk Sharing\thanks{I thank Christian Hellwig, Natalia Kovaleva, Eric Mengus, Harry Pei, Guillaume Plantin, Alp Simsek, Tong Su, Jean Tirole, Takuro Yamashita, and seminar participants at Toulouse School of Economics for helpful comments and discussions.}}
\author{Georgy Lukyanov\thanks{\raggedright Toulouse School of Economics, Universit\'e Toulouse Capitole, Toulouse, France. Email: \href{mailto:georgy.lukyanov@tse-fr.eu}{georgy.lukyanov@tse-fr.eu}.}}
\date{November 2026}
\begin{document}
\maketitle
\begin{abstract}
A credit mutual shares project returns among its borrowers. Outside lenders finance borrowers the mutual turns away. As the mutual admits riskier projects, safer borrowers must contribute more to the pool. Beyond a threshold they leave. Borrowers who already had credit then lose as a group. If the gains to newly financed borrowers outweigh this loss at the threshold, opening the market continues to raise ex ante welfare after exit. Near full capacity, the original members lose from outside access even when the mutual offers safer borrowers enough to stay.
\end{abstract}

\section{Introduction}
\label{sec:intro}

Consider a borrower whose mutual has turned down a profitable project because it already serves as many projects as it can handle. A commercial loan lets this borrower invest. A second borrower, already served by the mutual, may approach the same lender for a different reason. If her project has a high probability of success, she can obtain better terms on her own than under an agreement to share the proceeds with other borrowers. Which of these two uses of outside credit matters more to someone who does not yet know her project or her place in the queue?

We study a mutual that borrows on behalf of its members and insures the projects it admits. Members choose the agreement before they learn their types. For each dollar invested, safer projects yield more on average. Members therefore give these projects priority and admit riskier projects as capacity permits. Every admitted borrower receives the same insured benefit. We ask what happens if borrowers can subsequently contract with a lender on their own account, once project quality is public and before investment takes place. Outside lenders can supply insurance as well as debt. The mutual has already installed the capacity to service insurance contracts at a lower cost.

Suppose first that the agreement stands for the current lending cycle. A safer borrower stays if the mutual's advantage over her best outside contract covers the transfer she makes to other participants. When capacity is small, nearly everyone admitted has a safer project and the required transfer is small. Each new slot goes to a riskier project and raises that transfer. Eventually even a pool retaining every safer borrower offers too little. Beyond that capacity all safer borrowers finance their projects elsewhere. Below it, full retention can coexist with complete exit because no safer borrower wishes to join a pool from which all the others have departed. We use the equilibrium with the highest participation by safer borrowers to describe when they leave.

A riskier borrower already served loses the transfer from the safer projects. Another borrower of exactly the same type, originally denied a slot, now gets her project financed. She may even take over one of the slots vacated by the departing borrowers. At the capacity where the safer borrowers are just indifferent, the gains and losses of these two groups decide whether opening the market improves welfare after exit. If newly financed borrowers gain more than those already served lose, three thresholds follow in strict order. First the safer borrowers leave. At a higher capacity the riskier borrowers, taken together, lose from outside access. Only at a still higher capacity does access reduce the expected utility of the original membership. Under this condition all participation equilibria agree on whether access raises or lowers welfare relative to the closed agreement.

A vacant slot and a lost slot have different implications for this comparison. A relationship that can be reassigned lets another borrower use the mutual's cheaper insurance. A relationship that disappears with its former borrower leaves the next applicant reliant on the outside offer. Enough lost capacity can eliminate the interval over which access improves welfare after exit. If separation already yields a strict loss at that boundary, opening the market has opposite welfare effects in the two pure equilibria just below it.

The mutual can sometimes respond to competition. If it can turn applicants away, it can keep contributors by leaving some slots vacant. Borrowers who would previously have received mutual insurance bear the cost. Alternatively, the mutual can offer safer borrowers just enough to keep them and divide the balance among the others. Everyone the mutual serves then remains fully insured. Yet once capacity is sufficiently high, members would have preferred the closed arrangement before they knew their types. The transfer they give up among borrowers already served is worth more to them than access for the few borrowers still excluded.

One precedent for voluntary pooling comes from agricultural cooperatives. In \citet{Merel}, producers whose output differs in quality can share revenue for its insurance value, and the degree of pooling determines which groups can remain together. \citet{Abramitzky} finds that more productive members of Israeli kibbutzim, which pooled earnings, were more likely to leave. Collective wealth gave members a reason to stay. We study a lender whose service capacity limits the projects it can insure. An applicant denied a slot has a stake in the arrival of a competing lender even if she has never received a payment from the pool.

The vulnerability of a cross-subsidy to selective outside offers goes back at least to \citet{RS}. Risk is private information in their insurance market. Here a lender can condition both debt and insurance on a borrower's certified risk. In \citet{MW}, competition for profitable clients reduces the funds a nonprofit lender can devote to unprofitable clients. Here both types are profitable on their own. A mutual that loses its good risks can still finance other projects wherever it has service capacity. When contributors leave, the remaining projects lose a transfer. Whether opening the market benefits the original membership also depends on how many borrowers first obtain credit and on the insurance terms available outside.

\citet{BT97} study the allocation of borrowers between banks and markets. In \citet{BT}, borrower quality is public, bank service capacity is scarce, and banks invest in relationship lending. In \citet{DM}, lenders' information shapes the allocation of credit, including the use of scarce lending capacity across borrower segments. Here the return on a project is fixed. A marginal admission changes the amount payable to every incumbent by changing the average return of the insured pool. \citet{DFG} also study how lending contracts affect the welfare consequences of competition, in their case through rationing and the enforcement of loans.

The paper also relates to voluntary risk sharing. \citet{GR} examine arrangements from which individuals or groups can withdraw. \citet{ARR} show that public insurance can weaken private arrangements enough to reduce welfare. In our setting the outside offer arrives after the members have chosen their agreement. We compare the payments they can sustain for the current lending cycle with those they can sustain if the mutual revises its terms.

\section{An established sharing agreement}
\label{sec:model}

\subsection{Projects and service technologies}

There is a unit mass of borrowers, each with one project. Investment costs one and pays \(R\) on success and zero otherwise. The success probability is \(\theta_H\) for the safer type and \(\theta_L\) for the riskier type, where \(1>\theta_H>\theta_L>0\) and \(\theta_LR>1\). Each borrower initially has probability \(\mu\) of drawing H and probability \(m=1-\mu\) of drawing L. The realized masses are \(\mu\) and \(m\). Types and outputs are contractible and there is no hidden action. We take \(1/2\leq\mu<1\), so that capacity sufficient for all H is also sufficient for all L when slots can be reassigned.

A borrower's safe terminal endowment \(w>0\) cannot be pledged or used to finance investment. Without a loan she consumes \(w\) and obtains \(u_0=u(w)\), where \(u\) is twice continuously differentiable on \((0,\infty)\), with \(u'>0\) and \(u''<0\). In each admissible active pool the realized fraction of successful projects equals the pool's average success probability. This assumption leaves borrowers with individual project risk and the mutual with no aggregate output risk.

The borrowers jointly own a mutual with an installed capacity of \(k\in[\mu,1]\) insurance relationships. Enrolling a project takes one slot and costs \(c\geq0\) in operating expenses. In both regimes competitive risk-neutral investors finance projects and expenses at a unit gross return. The mutual can raise the money to fill every slot. What changes between regimes is whether a borrower can also approach a lender directly. Write
\begin{equation}
 s_i=\theta_iR-1,\qquad v_i=s_i-c,\qquad
 \Delta=v_H-v_L>0,\qquad \vb=\mu v_H+mv_L.
 \label{eq:surplus}
\end{equation}
Here \(s_i\) is expected project surplus, \(v_i\) is the surplus net of mutual service costs, and \(\vb\) is average net surplus in the population. The cost of installing \(k\) is sunk.

A borrower who leaves may take part of the mutual's capacity with her. If a mass \(x\) of H remains, the number of usable slots is
\begin{equation}
 K_\eta(x;k)=k-\eta(\mu-x),\qquad \eta\in[0,1].
 \label{eq:capacity}
\end{equation}
Each departure frees a slot at \(\eta=0\) and none at \(\eta=1\). For intermediate values it frees \(1-\eta\) slots per departing H. There is always room for every remaining H, since \(K_\eta(x;k)-x=(k-\mu)+(1-\eta)(\mu-x)\geq0\). Larger losses per departure could ration H themselves. We restrict attention to the range in \eqref{eq:capacity}.

An outside provider can issue debt at zero service cost or insure a project at cost \(f>c\). Either service is available at any scale. A group of H setting up a separate mutual would also pay \(f\) per insured project. Installed relationships can only be used within the existing mutual. Buying insurance from another provider cannot expand its supply of low-cost service during the cycle.

\subsection{The contract and its timing}

Suppose the mutual enrolls masses \(x\) of H and \(y\) of L. Its size is \(n=x+y>0\) and its fraction of successful projects is \(t=(x\theta_H+y\theta_L)/n\). Each successful borrower pays \(q\in[0,R]\). Once the mutual has repaid its investors, it divides what remains equally among enrolled borrowers, including those whose projects failed. The payment per borrower is
\begin{equation}
 p=tq-(1+c)\geq0.
 \label{eq:budget}
\end{equation}
An enrolled borrower consumes \(w+R-q+p\) on success and \(w+p\) on failure. Requiring \(p\geq0\) prevents the mutual from assessing members against their endowments. Borrowers whose projects are financed elsewhere or remain unfunded receive no distribution.

Before anyone knows her type, members choose a common payment and an admission rule for a world in which only the mutual can finance their projects. At this date everyone faces the same lottery, so the objective is the expected utility of one draw from the entire membership. Types then become public. Independent uniform priorities within each type break ties in admissions. We keep a borrower's priority fixed when comparing regimes.

After types become known and before borrowers pledge their projects, direct credit may unexpectedly become available. For this cycle, the mutual remains bound by the agreed repayment, eligibility rules, and order of admission. A borrower can apply at no cost and take the outside offer if rejected. A project can back only one contract. After investment, output is realized and the parties settle their claims.

The agreement fixes how to calculate the distribution. The amount paid depends on who enrolls. The operator can replace applicants who decline a slot but has no discretion to reserve slots, vary individual payments, or split the pool. Section~\ref{sec:adjustment} considers what happens if members give the operator either of the first two powers after direct credit becomes available.

A borrower who leaves takes this project elsewhere before funding. Previous loans and distributions are settled on the same terms in either regime. She may retain legal membership. What matters for the pool is where she finances this project.\footnote{Redeeming an individual capital account would require a payment to the departing member and a corresponding reduction in the mutual's assets. Equation~\eqref{eq:capacity} records service slots only.}

Fresno Madera Farm Credit, for example, bases its patronage distribution on eligible borrowing balances. Its loan prices also depend on risk, and its rules restrict eligibility for loans in nonaccrual or with write-offs \citep{FMFC}. Two borrowers with equal balances can therefore pay different net prices. The contract in \eqref{eq:budget} goes further by assuring each enrolled borrower the same distribution even if her own project fails. Appendix~\ref{app:contracts} varies this provision. The price and dividend choices in \citet{ESPrice} likewise give a reason to consider different terms for H and L.

\subsection{Outside finance and the interior case}

A competitive outside lender can advance one to type \(i\) against a success-contingent repayment of \(1/\theta_i\). Insuring the project instead entails taking its output, covering the loan and the cost \(f\), and paying a constant \(s_i-f\) to the borrower, provided this amount is nonnegative. Let \(D_i\) and \(b_i\) be expected utility and the certainty-equivalent benefit under debt. The best outside benefit \(a_i\) and the corresponding utility \(M_i\) satisfy
\begin{align}
 D_i&=\theta_i u\left(w+R-\frac{1}{\theta_i}\right)+(1-\theta_i)u_0,
 &b_i&=u^{-1}(D_i)-w,\notag\\
 a_i&=\max\{b_i,s_i-f\},&M_i&=u(w+a_i).
 \label{eq:outside}
\end{align}
The borrower earns strictly more than \(w\) on a successful debt-financed project, so \(b_i>0\). Equation~\eqref{eq:outside} also applies when \(s_i-f<0\) and the insured loan is unavailable.

Both products break even. The debt repayment is the smallest that covers the lender's advance. Once a provider incurs \(f\), it can offer a mean benefit of at most \(s_i-f\). A risk-averse borrower prefers a certain payment of that amount to any other allocation with the same mean. Partial coverage using this service and randomisation between the two products therefore leave the best outside value unchanged.

Throughout the baseline analysis we assume
\begin{equation}
 v_L>a_L>0,\qquad \vb<a_H<v_H.
 \label{eq:interior}
\end{equation}
An L borrower prefers even an all-L mutual to her best outside contract. An H borrower prefers an all-H mutual to that contract, but would borrow elsewhere if the mutual served the entire population on common terms. Let
\[
 u_L=u(w+v_L),\qquad u_H=u(w+v_H),\qquad
 r=\frac{a_H-v_L}{\Delta}\in(\mu,1).
\]
A proportion \(r\) of H in the pool gives each participant exactly \(a_H\). Appendix~\ref{app:regions} constructs economies satisfying \eqref{eq:interior} in which both types strictly prefer an insured outside loan to debt and \(c>0\).

\section{Participation and service capacity}
\label{sec:participation}

Suppose initially that only the mutual can finance a project. The members must decide how much of a successful project's return to collect, whose projects to admit, and whether to fill the last slot once the safest projects have all been admitted. Admitting an L at that point lowers each incumbent's payment but gives the additional borrower a project she could not otherwise finance.

\begin{lemma}
\label{lem:closed}
For any nonempty pool, the common contract that maximizes average expected utility has \(q=R\), uniquely. The closed agreement admits all \(\mu\) borrowers of type H and \(k-\mu\) of type L. Its payment to each admitted borrower, her utility, and expected utility over the original membership are respectively
\begin{equation}
 d(k)=v_L+\frac{\mu\Delta}{k},\qquad
 U(k)=u(w+d(k)),\qquad V^0(k)=kU(k)+(1-k)u_0.
 \label{eq:closed}
\end{equation}
The value \(V^0\) is strictly increasing in \(k\).
\end{lemma}

The members thus choose full insurance and full utilization in the closed regime. An individual L contributes less than the average but obtains a positive surplus from a slot. Before she knows which position she will occupy, a member wants that slot filled.

Now let borrowers obtain direct outside finance. An L prefers any available slot because its insured payment is at least \(v_L>a_L\). If a mass \(x\) of H enrolls, the mutual takes as many L as fit in the remaining slots. The resulting mass of L, total enrollment and payment per enrolled project are
\begin{align}
 y_\eta(x;k)&=\min\{m,k-\eta\mu-(1-\eta)x\},\notag\\
 n_\eta(x;k)&=x+y_\eta(x;k),\qquad
 d_\eta(x;k)=v_L+\Delta\frac{x}{n_\eta(x;k)}.
 \label{eq:pool}
\end{align}
The payment is defined for \(n_\eta>0\). An L turned away takes her outside contract. An H can obtain a slot whenever she applies. Taking the aggregate pool as given, she compares \(d_\eta(x;k)\) with \(a_H\). At every regular capacity \(k-\eta\mu>0\), the Nash conditions on the mass of H are
\[
 \begin{cases}
 d_\eta(0;k)\leq a_H,&x=0,\\
 d_\eta(x;k)=a_H,&0<x<\mu,\\
 d_\eta(\mu;k)\geq a_H,&x=\mu.
 \end{cases}
\]
We count projects financed by the mutual. Unsuccessful applications by borrowers who could equally go directly outside leave these masses unchanged. When more than one participation equilibrium exists, we first ask how many H the agreement can retain and take the equilibrium with the largest \(x\).

\begin{proposition}
\label{prop:participation}
In the equilibrium with the most H, all H stay for \(k\leq k^*\) and all H leave for \(k>k^*\), where
\begin{equation}
 k^*=\frac{\mu}{r}=\frac{\mu\Delta}{a_H-v_L}\in(\mu,1).
 \label{eq:participation}
\end{equation}
After exit the mutual serves \(\ell_\eta(k)=\min\{m,k-\eta\mu\}\) borrowers of type L. At regular capacities below \(k^*\), the equilibrium H masses are zero, one interior mass, and \(\mu\). At \(k^*\) only full retention and complete exit remain. Above \(k^*\) only complete exit survives.
\end{proposition}

The common payment rises as more H enroll. If \(k<k^*\), the payment in a pool with all H exceeds their outside value, while the payment in an all-L pool lies below it. At the interior equilibrium just enough H participate to make each indifferent. Appendix~\ref{app:participation} gives this mass and treats the sole case in which no service survives complete exit, \((k,\eta)=(\mu,1)\). There a returning H can reactivate her own slot. Even if a single borrower cannot restart service, full retention remains the equilibrium with the most H.

One can also derive \(k^*\) by comparing two dollar amounts. In a pool that retains every H, an H borrower transfers \(v_H-d(k)\) to the other participants. She is willing to give up at most \(v_H-a_H\) for access to mutual insurance. If she would buy insurance outside, this last amount is exactly the difference \(f-c\) in service costs. At \(k^*\) the two amounts coincide. Beyond it she can earn more by borrowing elsewhere even if every other H stays.

Changes in \(\eta\) have no effect at full retention since the mutual then has all its original slots. They do affect every calculation involving lost relationships. A higher \(\eta\) leaves fewer slots for L after H have begun to leave. Fewer H are then needed for the remaining pool to pay \(a_H\), so the interior equilibrium mass falls with \(\eta\). Complete exit also leaves fewer usable slots whenever L are rationed.

\section{Welfare and the distribution of gains}
\label{sec:welfare}

We evaluate direct credit from the date at which the members chose their agreement. For a given \(k\), we ask how much a randomly drawn member would gain if borrowers could finance projects outside. A member whose project was turned away and a member who later takes an outside loan both enter this calculation with their original weights.

With full retention (C), the mutual serves the same borrowers as before and previously excluded L obtain \(M_L\) instead of \(u_0\). With separation (S), each H obtains \(M_H\), each L using a mutual slot obtains \(u_L\), and each other L receives \(M_L\). Write
\begin{align}
 V^C(k)&=kU(k)+(1-k)M_L,\notag\\
 V^S_\eta(k)&=\mu M_H+\ell_\eta(k)u_L+[m-\ell_\eta(k)]M_L,\notag\\
 F_\eta(k)&=V^S_\eta(k)-V^0(k),\qquad G_\eta=F_\eta(k^*).
 \label{eq:values}
\end{align}
Since \(M_i>u_0\), all projects obtain finance in the open regime. Losing slots determines which L must use the outside product. The separated allocation is feasible as a comparison at every \(k\), including those at which full retention is selected. At the retention threshold \(U(k^*)=M_H\), hence
\begin{equation}
 G_\eta=(\mu-k^*)M_H+\ell_\eta(k^*)u_L
       +[m-\ell_\eta(k^*)]M_L-(1-k^*)u_0.
 \label{eq:welfare}
\end{equation}
At that capacity the H who leave neither gain nor lose. A previously admitted L forgoes a payment yielding \(M_H\) and receives the all-L payment yielding \(u_L\). A previously excluded L obtains either \(u_L\) or \(M_L\) in place of \(u_0\). The sign of \(G_\eta\) tells us which change is larger for the membership as a whole.

\begin{proposition}
\label{prop:welfare}
The welfare gain under separation, \(F_\eta(k)\), is continuous and strictly decreasing, and \(F_\eta(1)<0\). A unique \(k_W\in(k^*,1)\) with \(F_\eta(k_W)=0\) exists exactly when \(G_\eta>0\). In this case access raises the original cohort's welfare at every Nash allocation for \(k<k_W\), leaves it unchanged at \(k_W\), and lowers it for \(k>k_W\). Along the equilibrium with the most H, the gain to L on average changes sign once after exit, at \(k_L\), with
\[
 k^*<k_L<k_W<1.
\]
If \(G_\eta\leq0\), access raises welfare under full retention through \(k^*\) and lowers it above \(k^*\). If \(G_\eta<0\), retention and exit give opposite welfare rankings on an interval immediately below \(k^*\). If \(G_\eta=0\), all equilibria give a strict gain below \(k^*\), and at \(k^*\) separation gives equality while retention still gives a gain.
\end{proposition}

Increasing capacity helps the closed mutual more than it helps the separated allocation. The closed mutual can finance an otherwise unfunded project at every extra slot. After exit a slot moves at most one L from outside finance to the cheaper mutual service. Once all L fit, an additional slot is worthless to the separated pool. At \(k=1\) access finances no new projects at all.

When \(\eta=0\), all L fit after exit because \(\mu\geq m\). The effect of losing service on the welfare comparison at \(k^*\) can thus be written as
\[
 G_\eta=G_0-(u_L-M_L)[m-k^*+\eta\mu]_+,
 \qquad [z]_+=\max\{z,0\}.
\]
As long as the vacated slots would have stood empty, their loss costs nothing. Every further slot lost moves one L from \(u_L\) to \(M_L\). Appendix~\ref{app:regions} gives economies in which \(G_\eta>0\) even if every departure destroys a slot and others in which \(G_0>0>G_1\). Both cases occur when \(c>0\) and both types prefer insurance to debt outside.

Table~\ref{tab:groups} follows individual borrowers through the change in regime at \(k>k^*\), keeping their admission priorities fixed. Write
\[
 z=\ell_\eta(k)-(k-\mu)=\min\{1-k,(1-\eta)\mu\}\geq0.
\]
The bound of one lost slot per departing H ensures that all previously admitted L still fit. Of those previously excluded, \(z\) receive a mutual slot. The last two rows of the table report individual gains whenever the group has positive mass.

\begin{table}[ht]
\centering
\caption{Utility changes after selected H exit}
\label{tab:groups}
\small
\begin{tabular}{@{}p{0.34\textwidth}p{0.20\textwidth}p{0.26\textwidth}l@{}}
\toprule
Borrowers & Mass & Utility change & Sign\\
\midrule
H & \(\mu\) & \(M_H-U(k)\) & Positive\\
Previously admitted L & \(k-\mu\) & \(u_L-U(k)\) & Negative\\
Newly admitted L & \(z\) & \(u_L-u_0\) & Positive\\
L still outside the mutual & \(m-\ell_\eta(k)\) & \(M_L-u_0\) & Positive\\
\bottomrule
\end{tabular}
\end{table}

Which borrowers account for a positive welfare gain? Put \(E=M_L-u_0\), the gain from a first loan outside, and \(B=u_L-M_L\), the further gain from an all-L mutual slot. Summing the rows gives
\begin{align}
 F_\eta(k)={}&(1-k)E+zB-\mu(u_H-M_H)\notag\\
 &-\big[kU(k)-\mu u_H-(k-\mu)u_L\big]. \label{eq:welfaredecomp}
\end{align}
The first two terms accrue to borrowers previously turned away. The square bracket is nonnegative by concavity. It measures the value to the originally financed group of sharing across types at the same aggregate resources. That group also gives up the H borrowers' advantage from using the mutual's own service, \(\mu(u_H-M_H)>0\). Taken together, previously financed borrowers therefore lose at every separated allocation, although the individual H gain for \(k>k^*\). Any positive cohort gain must come from those originally excluded.

When \(G_\eta>0\) and \(k_L<k<k_W\), L lose on average but the gains to H still outweigh their loss at the original population weights. Those weights have an ex ante interpretation since each member draws the same lottery. To give a distributional comparison, let \(\alpha\) be the weight on average H utility and \(1-\alpha\) the weight on average L utility, leaving admission and finance unchanged. When \(G_\eta>0\), each weight \(0\leq\alpha\leq\mu\) gives its own welfare threshold between \(k_L\) and \(k_W\), and the sign on either side agrees across equilibria. A large enough weight on H reverses the welfare ranking even at full capacity. Appendix~\ref{app:weights} gives this weight.

At \(k=1\), only the two losses in \eqref{eq:welfaredecomp} remain. The first loss, \(u_H-M_H\) per H, reflects whichever outside contract H chooses. With insured outside finance H gives up the cost advantage \(f-c\). With debt she saves the mutual service cost but bears project risk. The second loss is the value of insurance against drawing the less productive type. The total number of projects and their gross output are the same in both regimes.

\section{Adjustment of the agreement}
\label{sec:adjustment}

Suppose now that, when outside credit becomes available, members can give their operator discretion over admissions or discretion over insured payments. We consider each power separately. The operator maximizes expected utility over the original membership. We continue to use the agreement those members would have chosen in the closed regime as the benchmark.

\subsection{Admission limits}

First let the operator announce a quota \(b\in[\mu,k]\) before applications. H keep priority and all admitted projects remain on the common contract. With \(x\) active H, enrollment cannot exceed the number of applicants, the announced quota, or the remaining physical capacity. Thus
\[
 n_b(x)=\min\{x+m,b,k-\eta\mu+\eta x\}.
\]
A quota limits admissions without changing the number of slots lost per departure. For each quota we evaluate the equilibrium with the most H. To retain every H, the quota must deliver a common benefit of at least \(a_H\) when they all stay.

For \(k>k^*\) the largest quota that pays this amount is \(k^*\). At this quota all admitted borrowers obtain \(M_H\) and those turned away obtain \(M_L\). Denote its value and the difference in favor of separation by
\begin{equation}
 R^*=k^*M_H+(1-k^*)M_L,\qquad
 Q_\eta(k)=V^S_\eta(k)-R^*.
 \label{eq:quota}
\end{equation}

\begin{proposition}
\label{prop:quota}
With H priority, a common contract, and selection of the equilibrium with the most H, the optimal quota gives value \(V^C(k)\) for \(k\leq k^*\) and
\[
 V^A_\eta(k)=\max\{R^*,V^S_\eta(k)\}\quad\text{for }k>k^*.
\]
For \(k>k^*\), an operator chooses separation when \(Q_\eta(k)>0\), retains all H at quota \(k^*\) when \(Q_\eta(k)<0\), and is indifferent at equality. For every \(G_\eta\) there is a unique \(k_W^A\in(k^*,1)\) at which optimal quota welfare equals \(V^0\). Access raises welfare below \(k_W^A\) and lowers it above. The switch to separation, when it occurs, can come before or after the welfare loss.
\end{proposition}

Who pays for retention through a quota? Of the L previously served, \(k^*-\mu\) keep their places and \(k-k^*\) must go outside. The \(1-k\) L originally turned away still obtain their first loans. The optimal quota trades the transfers preserved for the L who keep their places against the loss of mutual service for those displaced.

After complete exit, capacity determines how many L would receive the all-L mutual payment. The more slots would disappear, the less attractive this alternative becomes. The operator may thus keep all H by limiting admissions just above \(k^*\) and welcome their departure at a higher capacity. Appendix~\ref{app:quota} gives the cutoff and constructs examples with each ordering relative to \(k_W^A\).

\subsection{Differentiated insured benefits}

Alternatively, suppose the mutual can give the two types different insured payments. Consider allocations that keep every H, admit \(k-\mu\) borrowers of type L and fully insure each enrolled project. If the payments are \(z_H\) and \(z_L\), budget balance requires
\[
 \mu z_H+(k-\mu)z_L=\mu v_H+(k-\mu)v_L.
\]
Each admitted type must receive at least its outside benefit \(a_i\). We take H to stay at equality.

\begin{proposition}
\label{prop:transfers}
Among these retained, fully insured allocations, common payments \(d(k)\) maximize cohort welfare for \(k\leq k^*\). For \(k>k^*\) the optimal payments are
\begin{equation}
 z_H=a_H,\qquad
 z_L(k)=v_L+\frac{\mu(v_H-a_H)}{k-\mu}>v_L>a_L.
 \label{eq:transfers}
\end{equation}
These payments keep every H. There is a unique \(k_W^T\in(k^*,1)\) such that access improves ex ante welfare below \(k_W^T\) and reduces it above \(k_W^T\).
\end{proposition}

Above \(k^*\), the mutual can ask each H to contribute at most \(v_H-a_H\), precisely what that borrower would lose by taking her project elsewhere. The operator keeps this contribution and distributes it among the L who are admitted. No service slot disappears.

For \(k>k^*\), the cohort obtains
\[
 V^T(k)=\mu M_H+(k-\mu)u(w+z_L(k))+(1-k)M_L.
\]
The gain over the closed agreement is
\begin{align}
 V^T(k)-V^0(k)&=(1-k)(M_L-u_0)-\loss(k),\notag\\
 \loss(k)&=kU(k)-\mu M_H-(k-\mu)u(w+z_L(k)).
 \label{eq:transferloss}
\end{align}
The first term is the gain to people who previously had no loan. The second is the cost of paying different amounts to insured borrowers who would have received the same payment under the closed agreement. The sum of the payments to these borrowers is the same in both regimes. At \(k^*\) they can still share equally. Above \(k^*\) the utility cost of unequal sharing increases, while the number of members who need an outside loan to invest falls.

Thus access can reduce welfare with exactly the same people enrolled and fully insured as before. This comparison holds the set of funded mutual projects and the requirement of individual insurance fixed. When \(c>0\), giving up some insurance may also save service costs. Appendix~\ref{app:transfers} gives the optimal retained payments within the insured class, shows how to make staying strictly preferable to H at a small cost, and specifies feasible payments if a different pool forms.

Two changes in the contract show what is required for a mixed pool to survive. If \(f=c\), outside insurance pays H their entire net surplus \(v_H\), more than the common payment in any mixed pool. If the mutual distributes its residual only after success, an H in a mixed pool can do strictly better with fair debt outside. For a mutual with a cost advantage, the amount collected on success and the fraction of the residual payable after failure can both vary. Whenever L prefer an L-only mutual to their outside offer and H rank their outside offer between an H-only pool and the full mixed pool, the separated welfare gain still falls with capacity and is negative at full capacity. Thus a higher welfare threshold again exists exactly when the separated allocation improves welfare at the retention boundary. These inequalities hold in a neighborhood of full insurance whenever they are strict there. Appendix~\ref{app:contracts} proves the result.

\section{Borrower switching and the terms of membership}
\label{sec:implications}

For an empirical application, compare two pools with the same original membership and similar outside offers. Suppose one can service more insured loans. An expansion makes full retention impossible if the newly admitted projects dilute the common benefit enough to put it below the safer borrowers' outside offer. Expansion that preserves the mix of risks leaves the common benefit unchanged. Below the retention threshold, both staying and leaving can be equilibria. Operational resources do affect loan decisions \citep{Huang}, but neither total assets nor the number of loans outstanding measures this change in the composition of new admissions.

The amount to compare is the borrower's net payment across project outcomes. A smaller cash distribution may be offset by a lower loan charge or a larger retained claim. \citet{BJ} study lower-rate loans with patronage waivers, and \citet{ZMB} relate patronage decisions to capital management. In Fresno Madera's program a distribution based on the loan balance can coexist with a rate that reflects individual risk. It would take evidence on both terms and on eligibility after a loss to establish the insurance arrangement studied here.

Switching can occur for a different reason even when it is the safer borrowers who leave. In Rwanda, credit histories accumulated at cooperatives helped borrowers obtain commercial bank loans, with better terms for safer borrowers \citep{Agarwal}. In Uganda, entry by competing microfinance institutions lowered repayment rates and savings deposits at an incumbent without increasing its overall client dropout rate \citep{MDS}. \citet{FIM} find that commercial bank entry can improve loan terms both for borrowers who move and for clients who stay with their microfinance institution. These comparisons leave an open empirical question. Following an expansion of the incumbent's service capacity, do marginal admissions lower contributors' net benefits, and do the contributors whose benefits fall subsequently switch, holding their outside offers fixed?

Studies of competition between credit unions and banks measure several other outcomes. \citet{Feinberg01,Feinberg03} examine bank loan rates, and \citet{ESDynamic} study competition in local markets. \citet{CF} distinguish competition within cooperative networks from outside competition. \citet{GG} and \citet{Amendola} study how lender diversity and competition relate to bank profitability or performance. In our model average lending terms or profits leave out two relevant groups, those who have lost a transfer and those who previously had no loan. Tracking both groups in the original population would permit the welfare comparison in Table~\ref{tab:groups}.

\section{Conclusion}
\label{sec:conclusion}

Members who have yet to learn their project risks may want to share the proceeds of funded projects. Once risks are public, this requires a transfer a safer borrower can avoid by going elsewhere. A small mutual asks for little because most of its places go to the safer projects. At greater capacity each additional place goes to a riskier borrower, until even the agreement that retains every safer borrower pays too little to keep them.

Whether outside credit improves on the original agreement depends on the members originally turned away. As a group, the borrowers already financed lose when the safer borrowers depart. The excluded borrowers gain their first loans, and some can use the service slots released by departures. If their gains at the point of departure exceed the incumbents' loss, access continues to improve ex ante welfare over an interval in which the pool has already separated. Losing enough service capacity removes that interval.

A mutual that can revise its terms can retain its safer borrowers at every capacity. It can keep some riskier applicants out or give safer borrowers a larger insured payment. In the latter case all the originally served borrowers can still pool their project risks. Once the safer borrowers' participation constraint binds, members can still insure project failure in full, but they must bear more of the risk of drawing a low-return project.

\clearpage
\appendix
\section{Proofs}
\label{app:proofs}

\subsection{Outside finance and the closed mutual}
\label{app:closed}

Fair debt yields a positive benefit on success and zero on failure. Its mean is \(s_i\), so strict concavity gives \(0<b_i<s_i\). A safer borrower succeeds more often and keeps more output upon success. Thus \(b_H>b_L\), and \(s_H-f>s_L-f\) gives \(a_H>a_L\). Write \(\pi_i=s_i-b_i\) for the risk premium on debt. The two inequalities \(a_i<v_i\) hold precisely when \(c<\min\{\pi_H,\pi_L,f\}\).

\begin{proof}[Proof of Lemma~\ref{lem:closed}]
For a pool with success probability \(t\), mean consumption is \(w+tR-(1+c)\) at every feasible repayment. Setting \(q=R\) gives each borrower this amount in both states. The benefit is at least \(v_L>0\), so the contract is feasible and every admitted borrower accepts. Any \(q<R\) leaves consumption higher on success than on failure. Both states occur with positive probability, and strict Jensen gives uniqueness.

At fixed enrollment, replacing an L by an H adds \(\Delta\) to available resources per borrower replaced. Hence H receive priority. For an enrollment \(n\leq\mu\), cohort utility is \(nu_H+(1-n)u_0\), increasing in \(n\). Once all H have been admitted, let \(d=d(n)=v_L+\mu\Delta/n\). Each extra admission changes cohort utility at rate
\begin{equation}
 (V^0)'(n)=u(w+d)-u_0-(d-v_L)u'(w+d)
 \geq u_L-u_0>0.
 \label{eq:closedderivative}
\end{equation}
The inequality evaluates the tangent to \(u\) at \(w+d\) at the point \(w+v_L\). The optimum therefore fills the last slot as well.
\end{proof}

\subsection{Participation}
\label{app:participation}

\begin{proof}[Proof of Proposition~\ref{prop:participation}]
Put \(C=k-\eta\mu\). Suppose first that \(C>0\). Each H applicant fits and each L prefers admission to outside finance, even in an L-only pool. An available slot in an empty pool offers at least \(v_L\), so an empty pool cannot persist when capacity is positive. With H mass \(x\), the pool size is \(n(x)=\min\{x+m,C+\eta x\}\). Its H share increases strictly with \(x\). The two expressions for the share have derivatives \(m/(x+m)^2\) and \(C/(C+\eta x)^2\). They agree in value where the capacity constraint starts to bind.

At \(x=0\) the benefit is \(v_L<a_H\), so complete exit is always an equilibrium. At \(x=\mu\) the benefit is \(d(k)\), which covers \(a_H\) exactly when \(k\leq k^*\). An interior mass requires an H share of \(r\). When all L fit, the share is at most \(\mu<r\). The interior pool must therefore fill capacity. Solving \(x=r(C+\eta x)\) gives
\begin{equation}
 x^\dagger_\eta=\frac{rC}{1-\eta r},\qquad
 n^\dagger_\eta=\frac{C}{1-\eta r},\qquad
 y^\dagger_\eta=\frac{(1-r)C}{1-\eta r}.
 \label{eq:interiorpool}
\end{equation}
The denominator is positive, and \(x^\dagger_\eta<\mu\) exactly when \(k<k^*\). At such a root, \(n^\dagger_\eta=C+\eta x^\dagger_\eta\leq k\leq1\) and hence \(y^\dagger_\eta\leq1-r<m\). While the root is interior, its derivative with respect to \(\eta\) is \(r(rk-\mu)/(1-\eta r)^2<0\). Strict monotonicity of the share now gives the equilibrium H masses at all regular capacities:
\begin{equation}
 \E_\eta(k)=
 \begin{cases}
 \{0,x^\dagger_\eta,\mu\},&\mu\leq k<k^*,\\
 \{0,\mu\},& k=k^*,\\
 \{0\},& k>k^*.
 \end{cases}
 \label{eq:correspondence}
\end{equation}

The sole exception is \((k,\eta)=(\mu,1)\). Then \(n(x)=x\), and every positive pool pays \(v_H>a_H\). Full retention is the only positive equilibrium. At an empty pool we let a returning H use its own attached slot and receive \(v_H\), the limiting benefit in a positive pool. Zero then fails to be an equilibrium. If a single atomless borrower instead takes zero usable capacity as given, zero is also an equilibrium. Full retention has the greatest H mass under either convention. Finally, after complete exit the mass of L served is \(\min\{m,C\}\).
\end{proof}

\subsection{Welfare and the gains of L borrowers}
\label{app:welfare}

\begin{proof}[Proof of Proposition~\ref{prop:welfare}]
Write \(B=u_L-M_L>0\) and \(E=M_L-u_0>0\). After complete exit, an additional slot raises utility by \(B\) while some L remain outside and by zero once all L fit. In the closed mutual its marginal value is at least \(u_L-u_0>B\), by \eqref{eq:closedderivative}. Both value functions are continuous, including where the last L is admitted. Thus \(F_\eta\) falls strictly. All L fit at \(k=1\), and
\[
 V^S_\eta(1)=\mu M_H+m u_L
 <\mu u_H+m u_L<u(w+\vb)=V^0(1).
\]
The first inequality is H's preference for own-type mutual insurance. The second follows from strict concavity. It follows that \(F_\eta\) crosses zero strictly above \(k^*\) precisely when \(G_\eta>0\). Under retention, the gain equals \((1-k)E>0\) for every \(k\leq k^*\).

We next compare partial participation with the two endpoints. Write cohort utility at H mass \(x\) as
\begin{align*}
 W(x)&=\mu M_H+mM_L+x(M_L-M_H)+P(n(x),x),\\
 P(n,x)&=n\big[u(w+v_L+\Delta x/n)-M_L\big].
\end{align*}
The function \(P\) is jointly concave for \(n>0\). Indeed, when two pools are combined, their H shares are averaged with weights proportional to their enrolments, and concavity of \(u\) gives the inequality. At fixed \(x\), putting \(d=v_L+\Delta x/n\) gives
\[
 P_n=u(w+d)-M_L-(d-v_L)u'(w+d)\geq B>0.
\]
As \(n(x)\) is the minimum of two affine functions, it is concave. Joint concavity and the increase in \(n\) give concavity of \(W(x)\): an intermediate H mass has at least the enrollment obtained by interpolating between the two endpoint pools. Hence \(W(x)\geq\min\{W(0),W(\mu)\}\). At \((\mu,1)\) the same bound follows directly from \(W(x)=x u_H+(\mu-x)M_H+mM_L\).

If \(G_\eta>0\), both endpoint allocations gain through \(k^*\). The bound gives a strict gain for every interior equilibrium as well. Beyond \(k^*\) complete exit is unique. If \(G_\eta=0\), every equilibrium gains strictly below \(k^*\), while at \(k^*\) complete exit breaks even and retention gains. If \(G_\eta<0\), continuity gives a whole interval just below \(k^*\) where retention gains and exit loses. For either nonpositive value of \(G_\eta\), every capacity above \(k^*\) gives a loss.

For separation, let \(\Delta_H=M_H-U(k)\) and \(\Delta_L\) denote the changes in mean utility for H and L, respectively. Then
\begin{equation}
 m\Delta_L=\ell_\eta u_L+(m-\ell_\eta)M_L
 -(k-\mu)U(k)-(1-k)u_0,\qquad
 F_\eta=\mu\Delta_H+m\Delta_L.
 \label{eq:typegains}
\end{equation}
Since \(\Delta_H\) rises strictly with \(k\) and \(F_\eta\) falls, \(\Delta_L\) falls strictly too. At \(k=\mu\), \(m\Delta_L=mE+\ell_\eta(\mu)B>0\). At \(k=1\), \(\Delta_L=u_L-u(w+\vb)<0\). Its unique zero lies above \(k^*\) if \(G_\eta>0\), since \(m\Delta_L(k^*)=G_\eta\). At this zero H already gain, so cohort welfare is still higher than in the closed mutual.
\end{proof}

\subsection{When capacity loss changes the welfare ranking}
\label{app:regions}

Both signs of \(G_\eta\) can arise even when borrowers of either type would choose insurance over debt outside. Start with \(c=0\), fix positive surpluses \(s_L<s_H\), and choose \(0<f<\min\{\pi_H,\pi_L,s_L,\Delta\}\). Then \(a_i=s_i-f\), \(r=1-f/\Delta\), and \(k^*=\mu/r\). For fixed \(\mu\in(1/2,1)\), take \(f<m\Delta\) so that \(r>\mu\). At full slot loss,
\[
 G_1=(1-k^*)(M_L-u_0)-(k^*-\mu)(M_H-u_L)
 \longrightarrow m[u(w+s_L)-u_0]>0
 \quad\text{as } f\downarrow0.
\]
Since \(G_\eta\geq G_1\), sufficiently small positive \(f\) gives a gain at the exit boundary for every rate of slot loss.

For a sign reversal, set \(A=u(w+s_L)-u_0\), \(T=u(w+s_L-f)-u_0\), and \(Z=u(w+s_H-f)-u(w+s_L)\). Here \(A>T>0\) and \(Z>0\). At zero and full slot loss, respectively,
\[
 G_0=A-\frac{\mu}{r}[A+(1-r)Z],\qquad
 G_1=T-\frac{\mu}{r}[T+(1-r)Z].
\]
The two expressions vanish at
\[
 \mu_0=\frac{rA}{A+(1-r)Z},\qquad
 \mu_1=\frac{rT}{T+(1-r)Z}.
\]
We have \(\mu_1<\mu_0<r\), and both bounds approach one as \(f\downarrow0\). First choose \(f>0\) small enough for \(\mu_1>1/2\), then take \(\mu\in(\mu_1,\mu_0)\). Now \(G_0>0>G_1\). In both examples the choices of outside contract and all relevant inequalities are strict. They persist at sufficiently small positive \(c<f\).

\subsection{Admission quotas}
\label{app:quota}

\begin{proof}[Proof of Proposition~\ref{prop:quota}]
If all H apply, a quota \(b\in[\mu,k]\) yields enrollment \(b\) and common benefit \(d(b)\). For positive \(x\), the H share under the quota is
\[
 \frac{x}{n_b(x)}=\max\left\{\frac{x}{x+m},\frac{x}{b},\frac{x}{C+\eta x}\right\},\qquad C=k-\eta\mu.
\]
When \(C>0\), each share rises with \(x\). If \(b>k^*\), even the full H mass falls short of the share \(r\), so all H leave. At \(b\leq k^*\), full retention is an equilibrium and is selected. The exceptional case \(C=0\) has \(b=k=\mu<k^*\) and also selects retention.

At a retaining quota, the derivative of cohort utility with respect to \(b\) is
\[
 u(w+d(b))-M_L-(d(b)-v_L)u'(w+d(b))\geq B,
 \qquad B=u_L-M_L>0.
\]
Thus the operator chooses the largest retaining quota, \(\min\{k,k^*\}\), if it keeps H. Under separation, any inducing quota satisfies \(b>k^*>\mu\geq m\), so it never limits L enrollment. All such quotas yield \(V^S_\eta(k)\).

For \(k>k^*\), the difference between separation and capped retention reduces to \(Q_\eta(k)=B[\ell_\eta(k)-\ell^A]\), where
\begin{equation}
 \ell^A=\frac{(k^*-\mu)(M_H-M_L)}{B}>k^*-\mu,\qquad
 k^A_\eta=\eta\mu+\ell^A .
 \label{eq:policyboundary}
\end{equation}
The inequality follows from \(M_H>u_L\). If \(\ell^A>m\), capped retention is strictly better throughout. If \(\ell^A=m\), the two policies tie once all L can be served and retention is better before then. If \(\ell^A<m\), separation is strictly preferred at the threatened capacities \(k>k^A_\eta\), with a tie at equality. Here \(k^A_\eta\leq\mu+\ell^A<1\).

Let \(k^R\) solve \(V^0(k^R)=R^*\). Existence and uniqueness follow from the strict increase in \(V^0\) and the two endpoint comparisons. At \(k^*\), capped retention gains \((1-k^*)(M_L-u_0)>0\). At full capacity,
\[
 k^*a_H+(1-k^*)a_L
 =\mu v_H+(k^*-\mu)v_L+(1-k^*)a_L<\vb ,
\]
so Jensen gives \(R^*<V^0(1)\). Beyond \(k^*\), the optimal gain is the maximum of \(R^*-V^0(k)\) and \(F_\eta(k)\). Both are continuous and strictly decreasing. Their maximum has a positive right limit at \(k^*\) and is negative at one, giving the unique crossing \(k_W^A\).

At \(k^R\), the separated gain is \(Q_\eta(k^R)\). Hence \(k_W^A=k^R\) when \(Q_\eta(k^R)\leq0\), and \(k_W^A>k^R\) otherwise. If \(\ell^A<m\) and \(k^A_\eta>k^*\), the switch to exit occurs before, at, or after the welfare crossing according as \(k^A_\eta<k^R\), \(k^A_\eta=k^R\), or \(k^A_\eta>k^R\). If \(k^A_\eta\leq k^*\) in this case, the operator switches immediately above \(k^*\), while welfare still exceeds its closed value.
\end{proof}

To obtain both strict orderings, set \(c=0\), \(\mu=1/2\), fix \(s_H=s_L+\Delta>s_L>0\), and take \(0<f<\min\{\pi_H,\pi_L,s_L,m\Delta\}\). Both types buy insurance outside. As \(f\downarrow0\), \(k^*\to\mu\), \(k^*-\mu=\mu f/(\Delta-f)\), and \(B/f\to u'(w+s_L)\). Hence
\[
 \ell^A\longrightarrow L_0(s_L)
 =\frac{\mu[u(w+s_L+\Delta)-u(w+s_L)]}{\Delta u'(w+s_L)}
 \in(0,\mu).
\]
The upper bound is strict by concavity. With no slot loss, small \(f>0\) gives \(\ell^A<m\) and \(k^A_0<k^*\). Exit precedes the welfare loss.

At full slot loss the order can be reversed. In the limit \(f=0\), \(k^R\) solves
\[
 V^0(k)=\mu u(w+s_L+\Delta)+m u(w+s_L).
\]
As \(s_L\downarrow0\), this root tends to \(\mu\). At \(s_L=0\) the right side equals \(V^0(\mu)\), and \(V^0\) still rises strictly because the tangent inequality in \eqref{eq:closedderivative} is strict for \(d(n)>v_L\). Meanwhile \(L_0(s_L)\) tends to \(\mu[u(w+\Delta)-u(w)]/[\Delta u'(w)]>0\). First choose \(s_L>0\) small and then choose \(f>0\) small. Continuity gives \(k^R<\mu+\ell^A=k^A_1<1\), so retaining H remains optimal after welfare falls. Small increases in \(c\) and \(\mu\) preserve both strict orderings in their respective examples. The surpluses are feasible project primitives: choose \(R>1+s_H\) and put \(\theta_i=(1+s_i)/R\).

\subsection{Differentiated insured benefits}
\label{app:transfers}

\begin{proof}[Proof of Proposition~\ref{prop:transfers}]
At \(k\leq k^*\), common benefits satisfy participation and maximize utility by Jensen. For \(k>k^*\), making the full pool pay \(d(k)\) would give H less than \(a_H\). At any feasible \(z_H\geq a_H\), budget balance then gives \(z_L<z_H\). Raising \(z_H\) and financing it from L changes cohort utility at rate \(\mu[u'(w+z_H)-u'(w+z_L)]<0\). Thus \(z_H=a_H\), and the budget gives \eqref{eq:transfers}. With \(v_H>a_H\), the L constraint is slack. If H require a strict gain, any extra payment \(\varepsilon\in(0,v_H-a_H)\) leaves \(z_L>v_L\).

Above \(k^*\), the type-specific benefits satisfy \(a_H>d(k)>z_L(k)>v_L\) and average to \(d(k)\) among admitted borrowers. Hence the utility loss \(\loss(k)\) in \eqref{eq:transferloss} is positive. To see how it varies with capacity, put
\[
 g(z)=u(w+z)-(z-v_L)u'(w+z).
\]
For \(z>v_L\), \(g'(z)=-(z-v_L)u''(w+z)>0\). Differentiating \eqref{eq:transferloss} gives \(\loss'(k)=g(d(k))-g(z_L(k))>0\). At \(k^*\) both benefits equal \(a_H\) and the loss is zero. Thus the gain over the closed agreement is positive through \(k^*\), then declines strictly to \(-\loss(1)<0\).
\end{proof}

The proposed benefits can be financed for every pool a borrower might face. At the intended pool the operator pays the optimal benefits above. At any other positive pool it can pay each type \(v_i\), which balances the budget and satisfies both participation constraints. A single atomless deviation leaves the intended pool composition unchanged. H accept their outside value at equality, and every admitted L strictly prefers staying.

Allowing the operator to cap enrollment as well does not improve this retained allocation. Above \(k^*\), its marginal value of a filled slot is
\[
 (V^T)'(k)=g(z_L(k))-M_L\geq u_L-M_L>0.
\]
Below \(k^*\), the same conclusion follows from the derivative for retaining quotas in Appendix~\ref{app:quota}.

\section{Welfare weights and incomplete insurance}

\subsection{Weights on the two borrower types}
\label{app:weights}

Give weight \(\alpha\) to average H utility and \(1-\alpha\) to average L utility. For complete exit, \eqref{eq:typegains} gives
\begin{equation}
 \Delta\mathcal W^S_\alpha
 =\alpha\Delta_H+(1-\alpha)\Delta_L
 =\frac{\alpha}{\mu}F_\eta+\left(1-\frac{\alpha}{\mu}\right)\Delta_L .
 \label{eq:fixedweights}
\end{equation}
For \(\alpha\in[0,\mu]\) this is a convex combination of two strictly decreasing gains. At \(k^*\) it equals \((1-\alpha)G_\eta/m\), and at one it is negative. If \(G_\eta>0\), its unique zero runs from \(k_L\) at \(\alpha=0\) to \(k_W\) at \(\alpha=\mu\). At any such zero H strictly gain and L weakly lose. Giving H more weight therefore moves the zero strictly to the right.

For \(k\leq k^*\) and \(G_\eta>0\), Proposition~\ref{prop:welfare} gives a strict cohort gain at every equilibrium. H gain zero under retention and receive \(M_H\leq U(k)\) at every other equilibrium. L must consequently gain on average. Reducing the H weight below \(\mu\) preserves the gain. At \(k>k^*\), only separation is possible. If instead \(G_\eta\leq0\), \eqref{eq:fixedweights} is negative above \(k^*\), while the selected retained allocation still gains through \(k^*\).

For \(k>k^*\) with \(\Delta_L<0\), the minimum H weight for weak preference for access is \(-\Delta_L/(\Delta_H-\Delta_L)\). At one this threshold is
\begin{equation}
 \alpha^{\rm crit}(1)=\frac{u(w+\vb)-u_L}{M_H-u_L}\in(\mu,1).
 \label{eq:criticalweight}
\end{equation}
Indeed, at full capacity the cohort loses under population weights, while H strictly gain. Access is strictly preferred there exactly when \(\alpha>\alpha^{\rm crit}(1)\).

\subsection{Incomplete insurance and payments after failure}
\label{app:contracts}

Hold the repayment and residual rule fixed across the closed and open regimes. Let \(h_0=1+c\). Successful borrowers pay \(q\in[h_0/\theta_L,R]\) and receive a residual \(p\). Borrowers whose projects fail receive \(\gamma p\), for \(0\leq\gamma\leq1\). If the pool's success fraction is \(t\), the budget and payoffs are
\begin{align}
 A_\gamma(t)&=t+\gamma(1-t),&
 p(t)&=\frac{tq-h_0}{A_\gamma(t)},\notag\\
 c_S(t)&=w+R-q+p(t),& c_F(t)&=w+\gamma p(t),\notag\\
 U_i(t)&=\theta_i u(c_S(t))+(1-\theta_i)u(c_F(t)),&
 F(t)&=t u(c_S(t))+(1-t)u(c_F(t)).
 \label{eq:incomplete}
\end{align}
The lower bound on \(q\) makes the residual nonnegative even for an L-only pool. Here \(F(t)\) denotes the average utility of the pool. The conditions for an interior participation cutoff are
\begin{equation}
 L_{q\gamma}:=F(\theta_L)>M_L,\qquad
 U_H(\theta_H)>M_H>U_H(\theta_L+\mu\delta),
 \qquad \delta=\theta_H-\theta_L .
 \label{eq:incompleteconditions}
\end{equation}

Both types prefer a higher pool success fraction. Indeed, \(p_t=[q\gamma+(1-\gamma)h_0]/A_\gamma^2>0\) and \(c_S\geq c_F\). Thus \(U_i(t)\) and \(F(t)\) increase strictly. Moreover, \(p_{tt}=-2(1-\gamma)p_t/A_\gamma\), which gives
\begin{equation}
 F_{tt}=\frac{2\gamma p_t}{A_\gamma}[u'(c_S)-u'(c_F)]
 +p_t^2[t u''(c_S)+\gamma^2(1-t)u''(c_F)]<0.
 \label{eq:incompleteconcavity}
\end{equation}
The first term is nonpositive and the second is strictly negative.

The increase in \(F\) gives H priority in the closed mutual. When the pool contains only H, another admission yields \(U_H(\theta_H)-u_0>0\). Once all H are served, the marginal gain is \(F(t)-(t-\theta_L)F_t(t)-u_0\geq L_{q\gamma}-u_0>0\). Thus all slots are filled. In the open regime every L prefers admission. Under \eqref{eq:incompleteconditions} the equation \(U_H(\theta_L+r_{q\gamma}\delta)=M_H\) has a unique root \(r_{q\gamma}\in(\mu,1)\). The participation correspondence is \eqref{eq:correspondence}, with \(r_{q\gamma}\) replacing \(r\) and \(k^*_{q\gamma}=\mu/r_{q\gamma}\) replacing \(k^*\). At the empty-pool endpoint a returning H receives the H-only utility if able to reactivate its slot.

Write \(t(k)=\theta_L+\mu\delta/k\). The three cohort values are
\begin{align}
 V^0_{q\gamma}(k)&=kF(t(k))+(1-k)u_0,\notag\\
 V^C_{q\gamma}(k)&=kF(t(k))+(1-k)M_L,\notag\\
 V^S_{q\gamma,\eta}(k)&=\mu M_H+\ell_\eta(k)L_{q\gamma}
                         +[m-\ell_\eta(k)]M_L .
 \label{eq:incompletevalues}
\end{align}
The separated gain falls strictly with capacity. An extra slot in the closed mutual is worth at least \(L_{q\gamma}-u_0\), while after exit it is worth at most \(L_{q\gamma}-M_L\). At full capacity, put \(\bar t=\mu\theta_H+m\theta_L\). By H's preference for its own-type pool and strict concavity in \eqref{eq:incompleteconcavity},
\[
 V^S_{q\gamma,\eta}(1)=\mu M_H+mF(\theta_L)
 <\mu F(\theta_H)+mF(\theta_L)<F(\bar t)=V^0_{q\gamma}(1).
\]
Consequently, within \eqref{eq:incompleteconditions}, a welfare crossing strictly above the exit cutoff exists if and only if
\begin{equation}
 G_{q\gamma,\eta}:=
 V^S_{q\gamma,\eta}(k^*_{q\gamma})-V^0_{q\gamma}(k^*_{q\gamma})>0.
 \label{eq:incompletemargins}
\end{equation}
In evaluating \(V^0\) here the pool earns \(F(t(k^*_{q\gamma}))\) on average. It is H's own utility \(U_H\) that equals \(M_H\) at the cutoff.

The welfare comparison also holds for interior equilibria. In the argument of Appendix~\ref{app:welfare}, replace \(P(n,x)\) by \(n[F(\theta_L+\delta x/n)-M_L]\). This is jointly concave because \(F\) is concave, and its derivative in \(n\) is at least \(F(\theta_L)-M_L>0\). Welfare at an intermediate H mass is therefore at least the smaller endpoint value. At \((k,\eta)=(\mu,1)\) welfare is affine in \(x\) with slope \(U_H(\theta_H)-M_H>0\). When \(G_{q\gamma,\eta}>0\), both endpoints gain through \(k^*_{q\gamma}\), and above it only separation is possible.

At \((q,\gamma)=(R,1)\) the participation inequalities are strict. Since \(U_H\) has a strictly positive quality derivative, its indifference share varies continuously with the contract. So does \(G_{q\gamma,\eta}\), even if the pool lies at the kink in admissions. For each \(\eta\) with \(G_\eta>0\), the strict inequalities therefore persist at feasible contracts sufficiently close to full insurance.

At \(\gamma=0\), in contrast, the residual goes solely to successful borrowers. Substituting into \eqref{eq:incomplete} gives
\[
 c_S=w+R-\frac{1+c}{t},\qquad c_F=w.
\]
In a mixed pool \(t<\theta_H\), so \((1+c)/t>1/\theta_H\). An H borrower keeps strictly more on success with fair outside debt and consumes the same amount on failure. Thus no mixed pool can retain H under this rebate rule. With the baseline common benefit, setting \(f=c\) gives a different reason for exit: outside insurance alone gives H \(v_H\), which exceeds every mixed pool's common benefit.


\begin{thebibliography}{99}
\bibitem[Abramitzky(2008)]{Abramitzky}
Abramitzky, R. (2008). The limits of equality: Insights from the Israeli kibbutz. \emph{Quarterly Journal of Economics} 123(3), 1111--1159.

\bibitem[Agarwal et~al.(2023)]{Agarwal}
Agarwal, S., T. Kigabo, C. Minoiu, A. F. Presbitero, and A. F. Silva (2023). Serving the underserved: Microcredit as a pathway to commercial banks. \emph{Review of Economics and Statistics} 105(4), 780--797.

\bibitem[Amendola et~al.(2025)]{Amendola}
Amendola, A., C. Barra, M. Boccia, and A. Papaccio (2025). Diversity in banking: How does competition in the Italian banking market affect bank performance? \emph{Structural Change and Economic Dynamics} 73, 307--327.

\bibitem[Attanasio and Rios-Rull(2000)]{ARR}
Attanasio, O., and J.-V. Rios-Rull (2000). Consumption smoothing in island economies: Can public insurance reduce welfare? \emph{European Economic Review} 44(7), 1225--1258.

\bibitem[Boot and Thakor(1997)]{BT97}
Boot, A. W. A., and A. V. Thakor (1997). Financial system architecture. \emph{Review of Financial Studies} 10(3), 693--733.

\bibitem[Boot and Thakor(2000)]{BT}
Boot, A. W. A., and A. V. Thakor (2000). Can relationship banking survive competition? \emph{Journal of Finance} 55(2), 679--713.

\bibitem[Briggeman and Jorgensen(2009)]{BJ}
Briggeman, B., and Q. Jorgensen (2009). Farm credit member-borrowers' preferences for patronage payments. \emph{Agricultural Finance Review} 69(1), 88--97.

\bibitem[Coccorese and Ferri(2019)]{CF}
Coccorese, P., and G. Ferri (2019). Is competition among cooperative banks a negative sum game? \emph{Journal of Institutional Economics} 15(4), 673--694.

\bibitem[de Quidt et~al.(2018)]{DFG}
de Quidt, J., T. Fetzer, and M. Ghatak (2018). Market structure and borrower welfare in microfinance. \emph{Economic Journal} 128(610), 1019--1046.

\bibitem[Dell'Ariccia and Marquez(2004)]{DM}
Dell'Ariccia, G., and R. Marquez (2004). Information and bank credit allocation. \emph{Journal of Financial Economics} 72(1), 185--214.

\bibitem[Emmons and Schmid(2000)]{ESDynamic}
Emmons, W. R., and F. A. Schmid (2000). Banks vs. credit unions: Dynamic competition in local markets. Federal Reserve Bank of St. Louis Working Paper 2000-006A.

\bibitem[Emmons and Schmid(2002)]{ESPrice}
Emmons, W. R., and F. A. Schmid (2002). Pricing and dividend policies in open credit cooperatives. \emph{Journal of Institutional and Theoretical Economics} 158(2), 234--255.

\bibitem[Feinberg(2001)]{Feinberg01}
Feinberg, R. M. (2001). The competitive role of credit unions in small local financial services markets. \emph{Review of Economics and Statistics} 83(3), 560--563.

\bibitem[Feinberg(2003)]{Feinberg03}
Feinberg, R. M. (2003). The determinants of bank rates in local consumer lending markets: Comparing market and institution-level results. \emph{Southern Economic Journal} 70(1), 144--156.

\bibitem[Fresno Madera Farm Credit(2023)]{FMFC}
Fresno Madera Farm Credit (2023). \emph{Patronage Program Overview \& Frequently Asked Questions}. \url{https://www.fmfarmcredit.com/-/media/Files/Programs/2023-Patraonge-Definitions-and-FAQs-FINAL.ashx}.

\bibitem[Fu et~al.(2026)]{FIM}
Fu, J., V. Ioannidou, and M. Mishra (2026). Banking on competition: The spillover effects of bank entry into microfinance. \emph{Journal of Financial and Quantitative Analysis} 61(3), 1429--1458.

\bibitem[Garcia and Gonzaga(2024)]{GG}
Garcia, A. S., and A. L. M. Gonzaga (2024). How credit unions affect the profitability of Brazilian commercial banks? \emph{Quarterly Review of Economics and Finance} 93, 190--209.

\bibitem[Genicot and Ray(2003)]{GR}
Genicot, G., and D. Ray (2003). Group formation in risk-sharing arrangements. \emph{Review of Economic Studies} 70(1), 87--113.

\bibitem[Huang et~al.(2025)]{Huang}
Huang, B., J. Li, T.-C. Lin, M. Tai, and Z. Y. Zhou (2025). Attention constraints and financial inclusion. \emph{Journal of Financial and Quantitative Analysis} 60(4), 1727--1759.

\bibitem[McIntosh et~al.(2005)]{MDS}
McIntosh, C., A. de Janvry, and E. Sadoulet (2005). How rising competition among microfinance institutions affects incumbent lenders. \emph{Economic Journal} 115(506), 987--1004.

\bibitem[McIntosh and Wydick(2005)]{MW}
McIntosh, C., and B. Wydick (2005). Competition and microfinance. \emph{Journal of Development Economics} 78(2), 271--298.

\bibitem[M\'erel et~al.(2015)]{Merel}
M\'erel, P., T. L. Saitone, and R. J. Sexton (2015). Cooperative stability under stochastic quality and farmer heterogeneity. \emph{European Review of Agricultural Economics} 42(5), 765--795.

\bibitem[Rothschild and Stiglitz(1976)]{RS}
Rothschild, M., and J. Stiglitz (1976). Equilibrium in competitive insurance markets: An essay on the economics of imperfect information. \emph{Quarterly Journal of Economics} 90(4), 629--649.

\bibitem[Zhang et~al.(2013)]{ZMB}
Zhang, T., M. Mallory, and P. Barry (2013). Determinants of the patronage refund decision of Farm Credit System associations. \emph{Agricultural Finance Review} 73(1), 102--118.
\end{thebibliography}
\end{document}